Version: March 18, 2005

# Magnetic Vortex Resonance in Patterned Ferromagnetic Dots

V. Novosad, * F. Y. Fradin, P. E. Roy, K. Buchanan, K. Yu. Guslienko, and S. D. Bader

*Materials Science Division and Center for Nanoscale Materials, Argonne National Laboratory,*

*Argonne, IL 60439*

Abstract

We report a high-resolution experimental detection of the resonant behavior of magnetic vortices confined in small disk-shaped ferromagnetic dots. The samples are magnetically soft Fe-Ni disks of diameter 1.1 and 2.2 μm, and thickness 20 and 40 nm patterned via electron beam lithography onto microwave co-planar waveguides. The vortex excitation spectra were probed by a vector network analyzer operating in reflection mode, which records the derivative of the real and the imaginary impedance as a function of frequency. The spectra show well-defined resonance peaks in magnetic fields smaller than the characteristic vortex annihilation field. Resonances at 162 and 272 MHz were detected for 2.2 and 1.1 μm disks with thickness 40 nm, respectively. A resonance peak at 83 MHz was detected for 20-nm thick, 2-μm diameter disks. The resonance frequencies exhibit weak field dependence, and scale as a function of the dot geometrical aspect ratio. The measured frequencies are well described by micromagnetic and analytical calculations that rely only on known properties of the dots (such as the dot diameter, thickness, saturation magnetization, and exchange stiffness constant) without any adjustable parameters. We find that the observed resonance originates from the translational motion of the magnetic vortex core.

*) Author to whom correspondence should be addressed. Electronic address: novosad@anl.gov;

Advances in nanolithography in combination with thin film growth methods offer a unique opportunity to prepare a variety of novel laterally-confined nanostructured magnets.[1] Patterned nanomagnets are of great interest from both fundamental and applied viewpoints (see Refs. 1, 2 and references therein). They are promising candidates for ultrahigh density discrete recording media, magnetic random access memory cells, miniature magnetic field sensors, and spintronic logic devices. The magnetic behavior of spin assemblies is determined by the effective magnetic field acting on each individual spin in the system. The phase diagram of magnetically soft elements (*e.g.* for the case of the magnetocrystalline anisotropy being negligibly small) is governed by the competing contributions from the exchange between nearest neighbors and long range dipole-dipole (magnetostatic) energies.[3] The lowest total energy of lithographically patterned micron- and sub-micron disk-shaped particle arrays in remanence is usually a magnetic vortex state.[4] Both the static and dynamic properties of magnetic vortices have attracted much attention recently. The stability of magnetic vortices as a function of an external magnetic field, dot geometry, and long-range interdot magnetostatic interactions[4-8] is now well understood. The vortex magnetization distribution leads to a considerable modification of the nature of the spin excitations in comparison to those in the uniform (saturated) state. An excitation corresponding to spiral-like vortex core precession around the dot center was predicted[9] and later observed experimentally using a time-resolved Kerr effect technique.[10] This vortex core precession mode has a rather low eigenfrequency, <1 GHz, for dots with thicknesses much smaller than the diameter. The thermally populated higher frequency spin waves associated with the vortex ground state were first experimentally detected via Brillouin scattering in Ref. 11 and then investigated in more detail in Ref. 12 and calculated analytically and numerically in Refs. 12 and 13. Brillouin scattering, however does not permit the detection of eigenfrequencies in the sub-GHz range. Vortex core motion induced by pulsed magnetic fields has recently been detected by direct imaging with time-resolved magneto-optical[10,14] and photoemission electron microcopy techniques.[15] However, due to experimental limitations, the earlier data, provided only qualitative or semi-quantitative information about the low frequency region of the spin excitation spectrum.

The present work aims to complement our fundamental understanding of the magnetic vortices trapped in small ferromagnetic particles. We report herein the experimental detection of spin excitations corresponding to the translational mode of magnetic vortices, i.e. the low-frequency excitation mode corresponding to rotation of the vortex core (soliton part of magnetic vortex) around an equilibrium position, for patterned dots as a function of geometrical aspect ratio  The dynamic properties of the magnetic vortices are probed using a vector network analyzer in the microwave reflection mode, employing a coplanar waveguide.[16] This method provides frequency domain information similar to that obtained by ferromagnetic resonance (FMR) techniques; however, unlike FMR, a wide range of frequencies can be probed, and both the real and imaginary impedance measured. This is a particularly important for exploring vortex dynamics as the frequency of the vortex motion is generally too low to be detected using the standard FMR cavity techniques.  The experimentally measured frequencies are compared to micromagnetic and analytical calculations that rely only on the known properties, *i.e.* dot diameter, thickness, saturation magnetization ($M_s$), and the exchange stiffness constant, without any adjustable parameters.

Micrometer-sized permalloy ($Fe_{20}Ni_{80}$) disks were fabricated directly on top of microwave coplanar waveguides. A bilayer resist structure was used to facilitate the lift-off process.[17] Atomic Force Microscope images of the samples show that the dot diameters are 1.1 and 2.2 $\mu$m, slightly larger than the corresponding design values of 1 and 2 $\mu$m. The insert in Fig. 1 shows an optical micrograph of a portion of a sample with 2.2-$\mu$m-diameter dots. Separate MFM images and magneto-optical hysteresis loop measurements taken on a larger dot array (Fig. 1) reveal the presence of a magnetic vortex state in each individual particle and the accompanying characteristic magnetization reversal via nucleation, displacement and annihilation of magnetic vortices.[6] An Anritsu vector network analyzer (frequency range: 40 MHz - 20 GHz) was used in reflection mode to record the derivative of the real and the imaginary impedance as a function of frequency. Typically a 1-4 mW drive signal was applied, providing a sinusoidal in-plane excitation field of ~1 Oe. This configuration favors excitation of the translational mode of the magnetic vortex. Further, an electromagnet was used to apply a static in-plane magnetic field parallel to the waveguide (perpendicular to the dynamic field) and  field modulation (1 - 10 Oe) coils were used parallel to the static field, as is commonly used for cavity FMR measurements. Digital lock-in detection is synchronized to the field modulation signal. Measuring the derivatives of the absorption and the dispersion improves the signal-to-noise ratio and ensures that the recorded signal is magnetic in origin.  The waveguides consist of 250 nm of Cr/Au with a central trace width of 3 μm and a gap width of 1 μm. There are ~1,200 disks on each waveguide.

Figure 2 shows typical dispersion derivative spectra recorded for permalloy disks of diameters 1.1 and 2.2 μm and thickness 40 nm in a static in-plane magnetic field of 20 Oe. The frequencies of the resonances are 272 and 161 MHz, respectively. A resonance at 83 MHz was detected for 20-nm thick, 2-μm diameter disks. The linewidths are 16 MHz, 11 MHz and 2 MHz, respectively; the narrow linewidths indicate that the vortex resonance peaks were detected with high accuracy. Because of the phase-sensitive detection, the measurements in positive and negative fields of the same amplitude produce signals of opposite sign, and no signal can be detected in remanence at zero field. With increasing static applied field, the magnitude of the resonance line continues to increase until the vortex structure is suppressed. In remanence, the magnetic vortex is perfectly centered, whereas applying a magnetic field leads to a displacement of the vortex core. The disk is therefore subjected to an effective magnetic field equal to a superposition of the external and demagnetizing fields that remain unchanged during the reversible vortex core shift. This simplified argument suggests that there should be no field dependence of the resonance frequency. Experimentally, we find a rather small resonance line-shift (<~5%) that becomes noticeable when the magnetic field approaches the magnitude of the vortex annihilation field. This may be related to the deformation of the total vortex spin structure and the associated increase in the contribution of magnetic exchange energy. In general, the spin excitation spectrum of such a small ferromagnetic element is expected to be dominated by strong magnetostatic contributions defined by the dot geometry. Indeed, the width and the frequency of the resonance lines increase drastically with a decrease in particle diameter for a given thickness.

In order to understand the underlying physics we performed numerical micromagnetic and analytical calculations. The micromagnetic simulations used the Object Oriented Micromagnetic Framework (OOMMF).[18] Material parameters used were those typical for permalloy, with an exchange stiffness constant A=1.3 μerg/cm, damping parameter $\alpha$ = 0.01, and a magnetocrystalline anisotropy constant K=0. The saturation magnetization $M_s$=750 emu/cm$^3$, was deduced from separate SQUID and high-field FMR measurements of continuous (reference) films. Note that the default gyromagnetic ratio γ = 2.80 MHz/Oe (1.76*10$^{11}$ (Ts)$^{-1}$) in OOMMF was used both in the simulations and in extracting $M_s$ from the FMR. Modelling of the spin dynamics for the distinctive cases of "free" oscillation in zero applied field and "forced" oscillation by a time-alternating external field (of the form $H(t) = H_0 \cos(2\pi ft)$) were performed. The later aimed to mimic the conditions of our experiment in which the driving *ac* current serves to excite the spin structure. The resulting frequency, and the shape and half-width of the resonance line are in good agreement with the experiment.

Figure 3 shows a representative spectrum of the simulated translational vortex mode for a 40-nm-thick disk. The spectrum was obtained by first shifting the vortex from an initially centered vortex state in a static magnetic field of 50 Oe applied in the +*x*-direction. The system was then relaxed in zero field and the time evolution of the x-component of the average magnetization ($M_x$) was recorded. In order to obtain the frequency of the translational mode, a Fourier transform of the time evolution of $M_x$ was performed, yielding a frequency of 275 MHz. This frequency almost coincides with the corresponding experimental measurement. Note that these micromagnetic calculations rely only on known dot properties, such as the dot geometry, thickness, $M_s$, and the exchange constant A, and agreement is achieved without any adjustable parameters. The ac simulations yielded the same resonance frequency but at a higher computational expense. Due to limitations in computational resources, the dependence of the eigenfrequencies on the dot aspect ratio was calculated by varying the thickness of a 0.5-μm disk.

A realistic estimate of the magnetostatic contribution was used to calculate to the translational mode frequency analytically.[9] A rigid vortex model has been shown to describe the static (equilibrium) vortex shift in a magnetic field well.[6] This model assumes that the vortex spin structure moves in a magnetic field without deformation. However, such an approach overestimates the total energy (especially the magnetostatic contribution) and therefore is unable to realistically describe the dynamics of vortex movement. The magnetization distribution of a moving vortex is distinctively different from its static counterpart. In particular, the dynamic magnetization should satisfy strong pinning boundary conditions on the dot circumference.[19] Another approach to calculate the vortex dynamics is the "surface-charge-free" spin distribution model.[9] In this case, the magnetization configuration of an oscillating vortex is modified at each given moment of time so that the magnetic "charges" on the particle's lateral surfaces are eliminated. This leads to a lowering of the total energy due to reduced surface charges that is partially compensated by an increase in the exchange energy and volume magnetic charges. Comparison of the spin distributions predicted by these models with the micromagnetic simulations reveal that the second model does indeed provide a more realistic spin distribution. Interestingly, the micromagnetic spin distributions also show that the core is distorted when in motion, a feature not captured adequately by either model. Figure 4 shows a comparison of the vortex core profile ($M_z$ component) between the dynamic and static cases. The cross-sections XX' and YY' are through a static central vortex, and cross-sections AA' and BB' are through a vortex in steady-state motion under the influence of an alternating magnetic field driven at resonance for the translational mode. The 3D-plot shows a blow-up of the $M_z$-component in the vortex core with a contour plot projected onto the xy-plane.

The theoretical description of the vortex oscillations utilizes the effective equation for vortex collective coordinates.[9,20] We consider here the vortex translational mode, where the vortex center (a topological soliton) oscillates about its equilibrium position in a disk-shaped particle with thickness $L$ and radius $R$. This mode has the lowest frequency in the vortex excitation spectrum. A 2D magnetization distribution is assumed (independent of the *z*-coordinate along the dot thickness): $\mathbf{m}(\boldsymbol{\rho},t) = \mathbf{M}(\boldsymbol{\rho},t)/M_s$, $\mathbf{m}^2 = 1$ To describe the magnetization distribution we use the complex function $w(\zeta,\bar{\zeta}) = (m_x + im_y)/(1+m_z)$, where the complex variable is $\zeta = (x+iy)/R$. The function $w(\zeta,\bar{\zeta})$ is defined as $w(\zeta,\bar{\zeta}) = f(\zeta)$ if $|f(\zeta)|<1$ (within the vortex core) and $w(\zeta,\bar{\zeta}) = f(\zeta)/|f(\zeta)|$ if $|f(\zeta)| \geq 1$, where $f(\zeta)$ is an analytical function. In the case of magnetic vortex motion in a flat cylindrical dot, the function $f(\zeta) = (1/c)\left(i\zeta C + (a - \bar{a}\zeta^2)/2\right)$ is used, where $c=b/R$ is the normalized core radius, and $C = \pm 1$ describes the vortex chirality (counter-clockwise or clock-wise rotation of the static vector $\mathbf{m}$ in the dot plane. This model describes the case of zero side surface charges [$(\mathbf{m} \cdot \mathbf{n})_S = 0$] and predicts experimentally confirmed values of the translation mode eigenfrequency.[9] The function $f(\zeta)$ for the complex parameter $a$ ($|a|<1$) describes the magnetization distribution as a superposition of two vortices, one centered within the dot and the other is an image vortex located outside the dot. The function $f(\zeta)$ corresponds to the simplest dynamic magnetization distribution which satisfies strong pinning boundary conditions on the dot circumference $\rho$=$R$.[19] We consider the complex parameter $a = a' + ia''$ as a 2D vector $\mathbf{a} = (a_x, a_y)$ with components $a_x = a'$ and $a_y = a''$. The vector $\mathbf{a}$ corresponds to the volume-averaged magnetization of the shifted vortex $\langle \mathbf{m(r)} \rangle_V = \mathbf{a}/3$. It is perpendicular to the vortex core shift vector $\mathbf{s}$=$\mathbf{X}$/R and can be expressed as $\boldsymbol{a}$= -2C $\boldsymbol{z} \times \boldsymbol{s}$ for

*s*<<1. The case *a*=0 and s=0 corresponds to a centered vortex. *i.e.,* the oscillations of the averaged magnetization $\langle \mathbf{m}(\mathbf{r},t) \rangle_V$ are due to the vortex center position oscillations **X**(t). The magnetization distribution outside the vortex core described by the function $f(\zeta)$ is in good agreement with the micromagnetically simulated distribution. The core occupies a relatively small area of the dot, therefore the vortex motion is determined mainly by magnetostatic forces which is due to dynamic magnetic charges outside the vortex core. These charges (volume and side-surface) are properly accounted within the analytical model described above, i.e., we can use this model to find the vortex translation frequency.

The *z*-component of the vortex magnetization (vortex core profile) can be expressed as

$$m_z(x,y) = \frac{1 - \left| w(\zeta, \overline{\zeta}) \right|^2}{1 + \left| w(\zeta, \overline{\zeta}) \right|^2} \qquad . \tag{1}$$

For a small displacement of the vortex center from its equilibrium position (**X**=0 without a *dc* field) the vortex profile, in a coordinate system with the origin in the shifted vortex core center, is determined by the function:

$$\left| w(\zeta, \overline{\zeta}) \right|^2 = \frac{1}{c^2} \rho'^2 \left[ 1 - 2(\vec{\rho}' \cdot \vec{s}) \right], \qquad \vec{\rho}' = \vec{\rho} - \vec{s}, \qquad \text{s}<<1. \tag{2}$$

Note that the second term in the bracket corresponds to a deviation from the rigid vortex model. The dynamic vortex has an asymmetric profile along the radius vector **ρ** connecting the center of the dot and the center of the vortex (see Fig. 4, cross-section AA'). The vortex profile along the plane perpendicular to **ρ** in the point **ρ=s** (core position) is, however, symmetric, see cross-section BB" in Fig. 4. Equation (2) qualitatively corresponds to our micromagnetically calculated dynamic vortex profile (see the 3D plot in Fig. 4)

The eigenfrequency of the vortex translational mode in a cylindrical dot calculated within the side-charges-free model is given by:

$$\omega_0 = 2\gamma M_s \left[ 4\pi F_v(\beta) - 0.5(R_0 / R)^2 \right], \tag{3}$$

where $\gamma$ is the gyromagnetyic constant , $F_v(\beta) = \int_0^\infty dt t^{-1} f(\beta t) I^2(t)$ is a function of the dot aspect ratio $\beta = L/R$, $f(x) = 1 - (1 - \exp(-x))/x$, and $R_0 = (2A/M_s^2)^{1/2}$ is the exchange length.

Using Eq. (3) one finds the translational mode frequencies of 310 and 161 MHz for dots of 40-nm thickness and 1.1- and 2.2-μm diameter, respectively, and 89 MHz for L=20 nm, R=2 μm dots, which are in good agreement with the experiment (see Fig. 5).[21] The vortex core rotates in counter-clockwise direction, if we assume $m_z(0) = +1$ (the frequency (3) is positive). Note that discrepancies in the analytical calculations are expected for higher aspect ratio dots because of the increasing non-uniformity of the magnetization distribution along the *z* axis. By performing micromagnetic calculations for disks with variable thickness, we were able to confirm previously reported scaling of the frequencies with the dot aspect ratio $\beta$=*L/R*.[9] This means that the vortex translation frequency in circular micron-size dots is determined mainly by magnetostatic interactions. We did not observe either experimentally or numerically a second vortex translation mode predicted to be in the same frequency range.[22] This is, however, to be expected since in both our experiment and micromagnetic simulations the applied field was an in-plane ac magnetic field along one direction. The second mode corresponds to another direction of the vortex core rotation (negative frequency) and might therefore be excited by applying in-plane rotating magnetic fields. We note, however, that our Eq. (3) explains with reasonable accuracy the time-resolved Kerr measurements by Park *et al.*[10] (see Fig. 5), where also only one vortex mode was detected.

In summary, we have experimentally detected the resonant behavior of magnetic spin vortices trapped in small disk-shaped ferromagnetic dots using a microwave absorption technique. The resonance frequency shows weak field dependence, and scales as a function of the dot geometrical aspect ratio. The measured frequencies are in good agreement with micromagnetic and analytical calculations that rely only on the known dot properties, such as dot diameter and thickness, the saturation magnetization, and the exchange stiffness constant, without any adjustable parameters. The observed resonance peak originates from the translational mode of the magnetic vortex core motion around the equilibrium position.

We acknowledge stimulating discussions with Y. Otani and M. Grimsditch. K. B. acknowledges a postdoctoral fellowship from NSERC Canada. The work was supported by the US Department of Energy, BES Materials Sciences under contract W-31-109-ENG-38.

**Figure captions**
 to the manuscript by V. Novosad et al.:

Fig. 1) MFM images reveal the presence of magnetic vortices in lithographically patterned permalloy dots of thickness L= 40 nm and diameter 2R=1 μm.

Fig. 2) Frequency dependence of the dispersion derivative measured for three permalloy disk samples with a) L=20 nm, 2R= 2.0 μm, b) L=40 nm, 2R = 2.2 μm, and c) L=40, 2R=1.1 μm. The resonances at 83, 162 and 272 MHz are in agreement with the eigenfrequencies of the collective spin excitations simulated micromagnetically and analytically.

Fig. 3) Micromagnetically simulated spectrum of low frequency spin excitations in 40-nm thick and 1.1-μm diameter dots. The insert shows damped oscillations of the magnetization due to spiral-like vortex core motion around the center of the dot.

Fig. 4) Comparison of the vortex core profile ($M_z$ component) between dynamic and static cases. Cross-sections XX' and YY' are through a static central vortex and cross-sections AA´and BB' are through a vortex in steady-state motion under the influence of an alternating magnetic field driven at resonance for the translational mode. The 3D-plot shows a blowup of the $M_z$-component in the vortex core with a contour-plot projected on the *xy*-plane.

Fig. 5) Comparison of the present experimental data (markers), micromagnetic calculations (solid line), analytical theory (dotted line), and experimental data for the eigenfrequency of the vortex translational mode *vs.* the dot aspect ratio β=L/R.

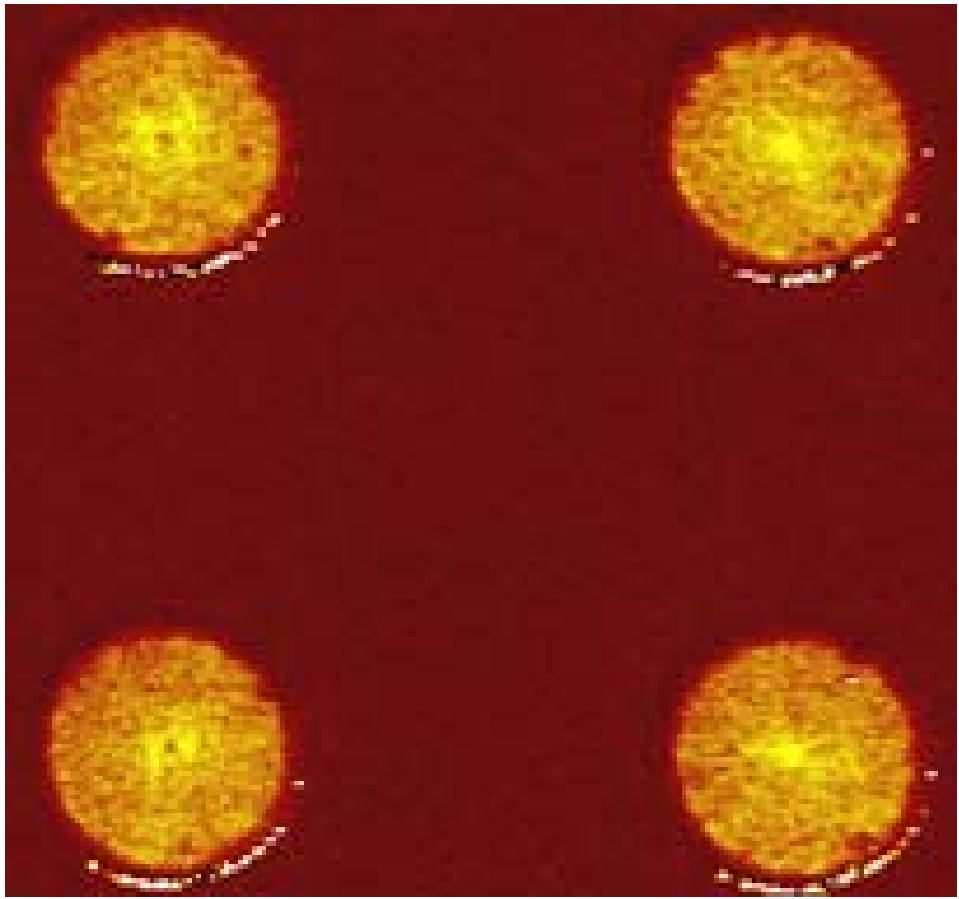

Fig. 1

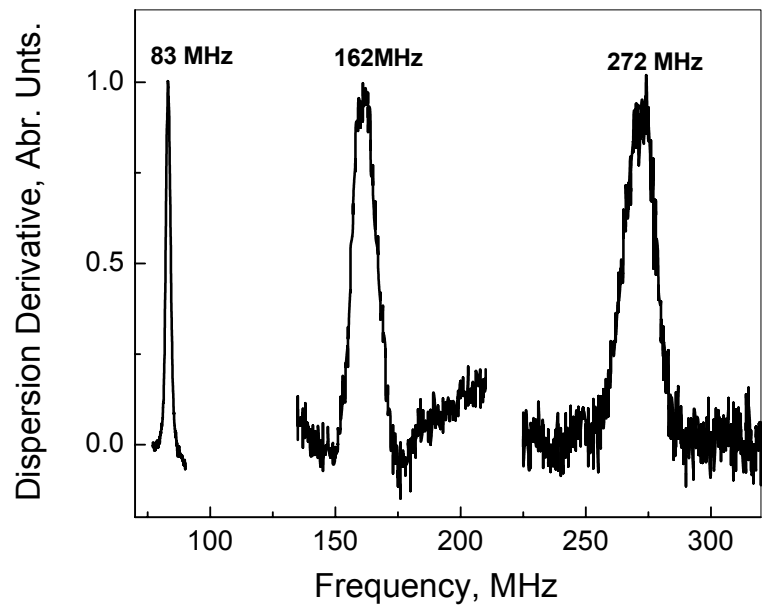

Fig. 2

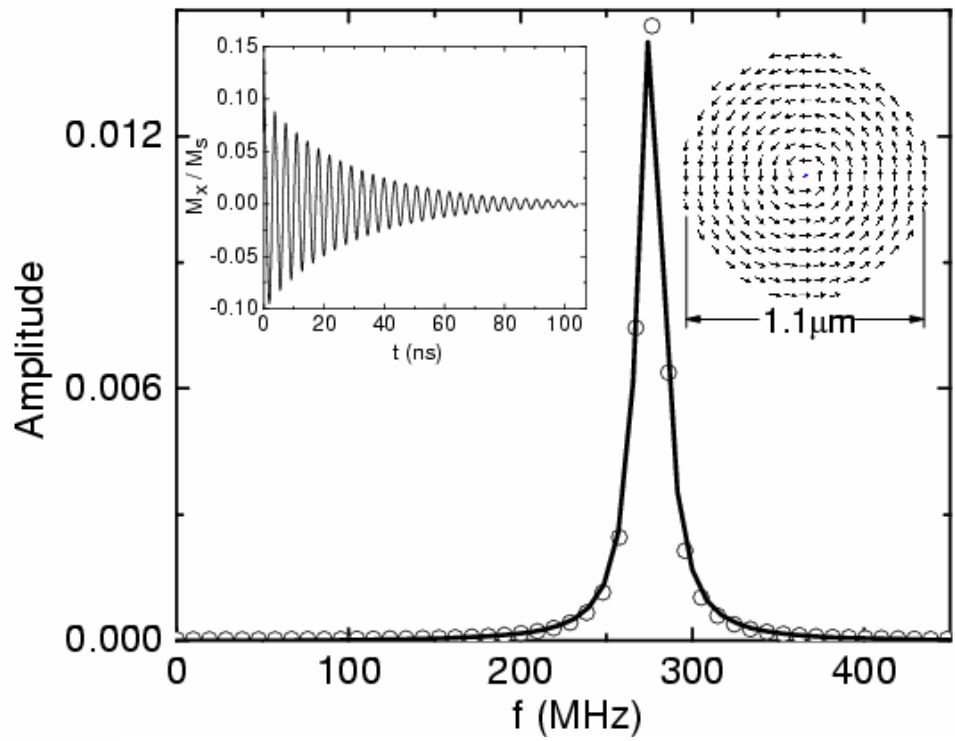

Fig. 3

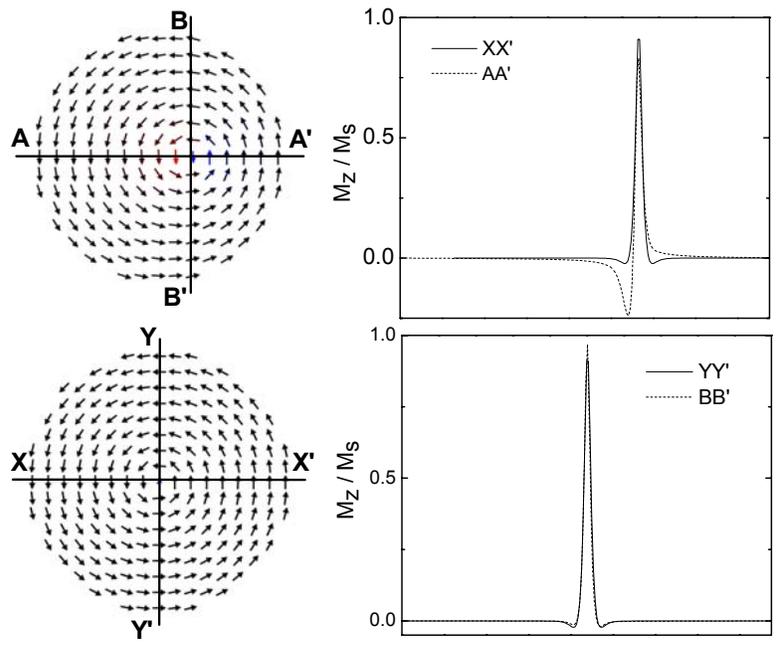

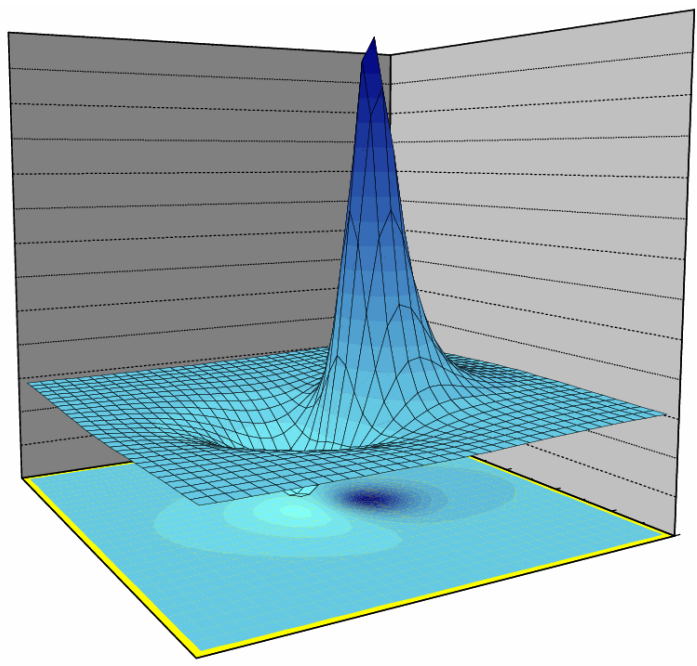

Fig. 4

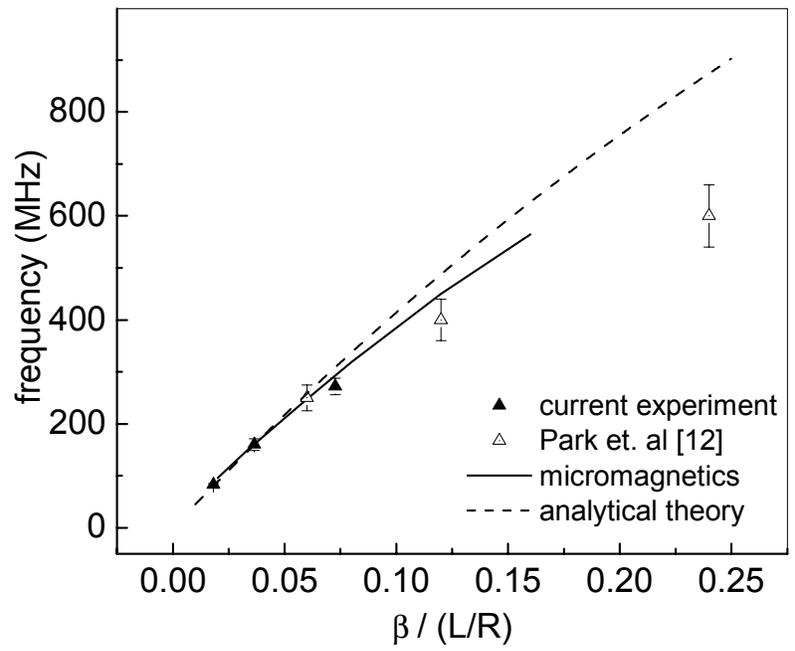